# Coexistence of ferroelectric and ferrielectric phases in ultrathin antiferroelectric PbZrO$_3$ thin films

Ying Liu[1,2,*], Ranming Niu[2], Roger Uriach[1], David Pesquera[1], José Manuel Caicedo Roque[1], José Santiso[1], Julie M Cairney[2], Xiaozhou Liao[2], Jordi Arbiol[1], Gustau Catalan[1,3,*]

[1]Catalan Institute of Nanoscience and Nanotechnology (ICN2), Campus Universitat Autonoma de Barcelona, Bellaterra 08193, Catalonia, Spain

[2]School of Aerospace, Mechanical and Mechatronic Engineering, The University of Sydney, Sydney, NSW 2008, Australia.

[3]Institut Català de Recerca i Estudis Avançats (ICREA), Pg. Lluís Companys 23, 08010 Barcelona, Catalunya.

**Correspondence to:**

Dr Ying Liu; Email: ying.liu6@sydney.edu.au; ORCID: 0000-0002-1262-2590

Prof Gustau Catalan; E-mail: gustau.catalan@icn2.cat; ORCID: 0000-0003-0214-4828

## Abstract

Whereas ferroelectricity may vanish in ultra-thin ferroelectric films, it is expected to emerge in ultra-thin anti-ferroelectric films, sparking people's interest in using antiferroelectric materials as an alternative to ferroelectric ones for high-density data storage applications. Lead Zirconate (PbZrO$_3$) is considered the prototype material for antiferroelectricity, and indeed previous studies indicated that nanoscale PbZrO$_3$ films exhibit ferroelectricity. The understanding of such phenomena from the microstructure aspect is crucial but still lacking. In this study, we fabricated a PbZrO$_3$ film with thicknesses varying from 5 nm to 80 nm. Using Piezoresponse Force Microscopy, we discovered the film displayed a transition from antiferroelectric behaviour in the thicker areas to ferroelectric behaviour in the thinner ones, with a critical thickness between 10 and 15 nm. In this critical thickness range, a 12 nm PZO thin film was chosen for further study using aberration-corrected scanning transmission electron microscopy. The investigation showed that the film comprises both ferroelectric and ferrielectric phases. The ferroelectric phase is characterized by polarisation along the [011]$_{pc}$ projection direction. The positions of Pb, Zr, and O were determined using the integrated differential phase contrast method. This allowed us to ascertain that the ferroelectric PbZrO$_3$ unit cell is half the size of that in the antiferroelectric phase on the *ab* plane. The observed unit cell is different from the electric field-induced ferroelectric rhombohedral phases. Additionally, we identified a ferrielectric phase with a unique up-up-zero-zero (↑↑··) dipole configuration. The finding is crucial for understanding the performance of ultrathin antiferroelectric thin films and the subsequent design and development of antiferroelectric devices.

# INTRODUCTION

Antiferroelectric (AFE) materials feature antiparallel electric dipoles that can be realigned into a parallel arrangement when subjected to a sufficiently high electric field.[1-4] This realignment can be viewed as a field-induced phase transition from AFE to ferroelectric (FE), a process that is marked by significant charge storage, volume expansion, and a change in temperature (electrocaloric effect).[5-8] Such properties make these materials suitable for use in a variety of electronic devices, including capacitors with high energy/power density, actuators capable of large strains, solid-state refrigeration, and advanced thermal switches.[6, 9-12] For most applications, AFE materials are best utilized when integrated into electronic devices as thin films. Moreover, the demand for smaller, higher-performance electronic devices that can be actuated with smaller voltages has led to the preparation of even thinner films. Yet, as antiferroelectric films are made thinner, it is often observed that their ground state evolves from antipolar to polar.[13] This represents both a challenge for applications relying on antiferroelectric switching, and an opportunity for applications, such as memory devices and ferroelectric tunnel junctions, where very thin ferroelectrics are desirable. In either case, identifying the critical thickness for the AFE-FE transition and determining how the structure evolves from one to the other (whether abruptly or through an intermediate range of phase coexistence), is essential.

$PbZrO_3$ (PZO) was the first antiferroelectric material discovered and is considered the archetype.[2] It is distinguished by an up-up-down-down (↑↑↓↓) dipole pattern within a unit cell.[3] In antiferroelectric PZO thin films or even in single crystals, various ferrielectric phases, which show antiparallel but uncompensated electric dipoles, have been identified to exist.[14-21] When the thickness of PZO thin films is reduced to the nanometer scale, they are reported to transition to a ferroelectric rhombohedral phase.[13] Since ferroelectric materials often suffer from reduced or completely lost polarisation at the heterointerface due to depolarisation effects,[22, 23] using antiferroelectric materials instead of ferroelectric ones in ultra-high-density information storage could offer a promising alternative. However, these ultrathin antiferroelectric films have not been extensively explored, and their ferroelectric properties have yet to be verified from a structural standpoint. The lack of this understanding has impeded their use in electronic devices.

In this work, we have fabricated a PZO thin film with thicknesses ranging from 5 to 80 nm and an ultrathin PZO film of about 12 nm. Combining X-ray diffraction, scanning probe microscopy and aberration-corrected transmission electron microscopy, we investigated the ferroelectricity and the phases in PZO epitaxial films, including one with a thickness gradient such that, within the same sample, the thick side is antiferroelectric and the thin end is ferroelectric.

# MATERIALS AND METHODS

*Thin film fabrication*: Using a Pulsed Laser Deposition (PLD) system with a KrF excimer laser (COMPex 102, Lambda Physik, λ = 248 nm), PZO thin films of about 12 nm were grown on $SrTiO_3$ (STO) (001) substrates (CrysTec GmbH) covered by $SrRuO_3$ (SRO) bottom electrodes. Before deposition, the base pressure of the chamber was brought down to $3.0 \times 10^{-3}$ mTorr. Both PZO and SRO layers were deposited under an oxygen pressure of 100 mTorr, with laser fluence of 2.0 – 2.5 J/cm² at a laser repetition rate of 2 Hz. The target-to-substrate distance was kept at 5 cm. The substrate temperature was set at 635°C when depositing the SRO layer, whereas the temperature was changed to 550 °C when growing the PZO layer. After deposition, the films were cooled down to room temperature at 5 °C/min. When depositing PZO thin films with a gradient in thickness, we employed a hard mask capable of precise positional control to shield certain areas of the substrate. Initially, the mask is positioned away from the substrate area, rendering it inactive. It is then incrementally moved to begin covering the substrate, with its coverage gradually extended until it completely shields the substrate as the deposition process progresses, culminating in the completion of the film deposition.

*X-ray diffraction (XRD) measurements*: High resolution $\theta$–$2\theta$ scans and reciprocal space maps were measured using a four-circle diffractometer equipped with 2×Ge(220) monochromator (Malvern PANalytical X'pert Pro MRD, Cu K$\alpha_1$, $\lambda$=1.5406 Å). The local measurements at different X-position across the sample were made by using a narrow divergence slit of 1/4º. Reciprocal space maps (RSM) were measured in the region where pseudocubic ($\bar{1}$03) of SRO and PZO films and STO substrate coexist.

*Scanning Probe Microscopy investigation (SPM):* Atomic Force Microscopy (AFM) and Piezoresponse Force Microscopy (PFM) experiments were carried out using a SPM system (MFP-3D Classic, Asylum Research, Santa Barbara, CA, USA) in tapping and dual-AC resonance tracking mode, respectively. The PPP-EFM probe (Nanosensors, Neuchâtel, Switzerland) was used for acquiring topographic images and phase and amplitude curves.

*Transmission Electron Microscopy (TEM) investigation*: Cross-sectional TEM specimens of PZO thin films oriented along pseudocubic [100] direction were prepared by slicing, gluing, polishing, and finally ion milling. An FEI Tecnai F20 Transmission Electron Microscope (operated at 200 kV) was employed for the low magnification morphology observation of the thin film. Atomic-scale scanning transmission electron microscopy high angle annular dark field (STEM-HAADF) and integrated differential phase contrast (iDPC) images were acquired using a Thermo Fisher Themis-Z Double-corrected 60–300 kV S/TEM. The convergence and collection angles under the STEM-HAADF mode are 17.9 mrad and 50 – 200 mrad, respectively, while those for iDPC images are 17.9 mrad and 9 – 35 mrad, respectively. The point resolution of Themis-Z under the STEM mode is around 0.6 Å (operated at 300 kV). The in-plane strain was analyzed using the Geometric Phase Analysis (GPA).[24-27] A Python library "Atomap"[28] was used to extract atom positions from atomic resolution STEM-HAADF and STEM-iDPC images by fitting 2D Gaussian functions to every atomic column in STEM-HAADF images. The extracted atomic positions were employed to determine Pb displacements with respect to their four nearest Zr ($\delta_{Pb}$).

## RESULTS AND DISCUSSION

Figure 1(A) shows the XRD $\theta$ – $2\theta$ scan for a PZO thin film whose thickness changes from 5 nm to 80 nm with a gradient of ∼7.5 nm/mm. Thickness at different positions was estimated from the FWHM of the PZO film reflections in the 2theta scan by applying the Scherrer equation, assuming the instrumental width from the substrate (002) peak. Two peaks for PZO can be seen in the 80 nm thick PZO film: one for the antiferroelectric orthorhombic (004)$_O$ (marked by a red arrow) and the other for (240)$_O$ (marked by a black arrow); both (004)$_O$ and (240)$_O$ correspond to the pseudo-cubic (002) peak of PZO. The

(004)$_O$ and (240)$_O$ correspond to the *c*-axis of PZO parallel (*c*-parallel) and perpendicular (*c*-perpendicular) to the film surface, respectively. As the thickness decreases, the two peaks move closer and merge into a single wide peak, which is supposed to be a superposition of antiferroelectric and ferroelectric phase. For ultrathin PZO films, an AFE to FE transition was reported in a previous study.[13] Then the out-of-plane lattice constant of PZO was calculated from the XRD data, with the findings displayed in Figure 1(B). Here, black squares represent the (240)$_O$ peaks, and red circles represent the (004)$_O$ peaks. For the regions of the film that are estimated to be 12.5 nm and 9 nm thick, the two peaks combine into one. This chart helps us identify a critical thickness between 10 – 15 nm where the AFE to FE phase change is likely to occur. RSM tests were also carried out to further support the XRD findings shown in Figure 1(A). Three locations were chosen from the gradient PZO thin film with thicknesses of 80 nm, 45 nm, and 9 nm. The outcomes are presented in Figures 1(C) to (E). For the 80 nm and 45 nm PZO areas, two distinct peaks are visible, but for the 9 nm film, only one peak is evident. It further confirms the AFE to FE phase transition.

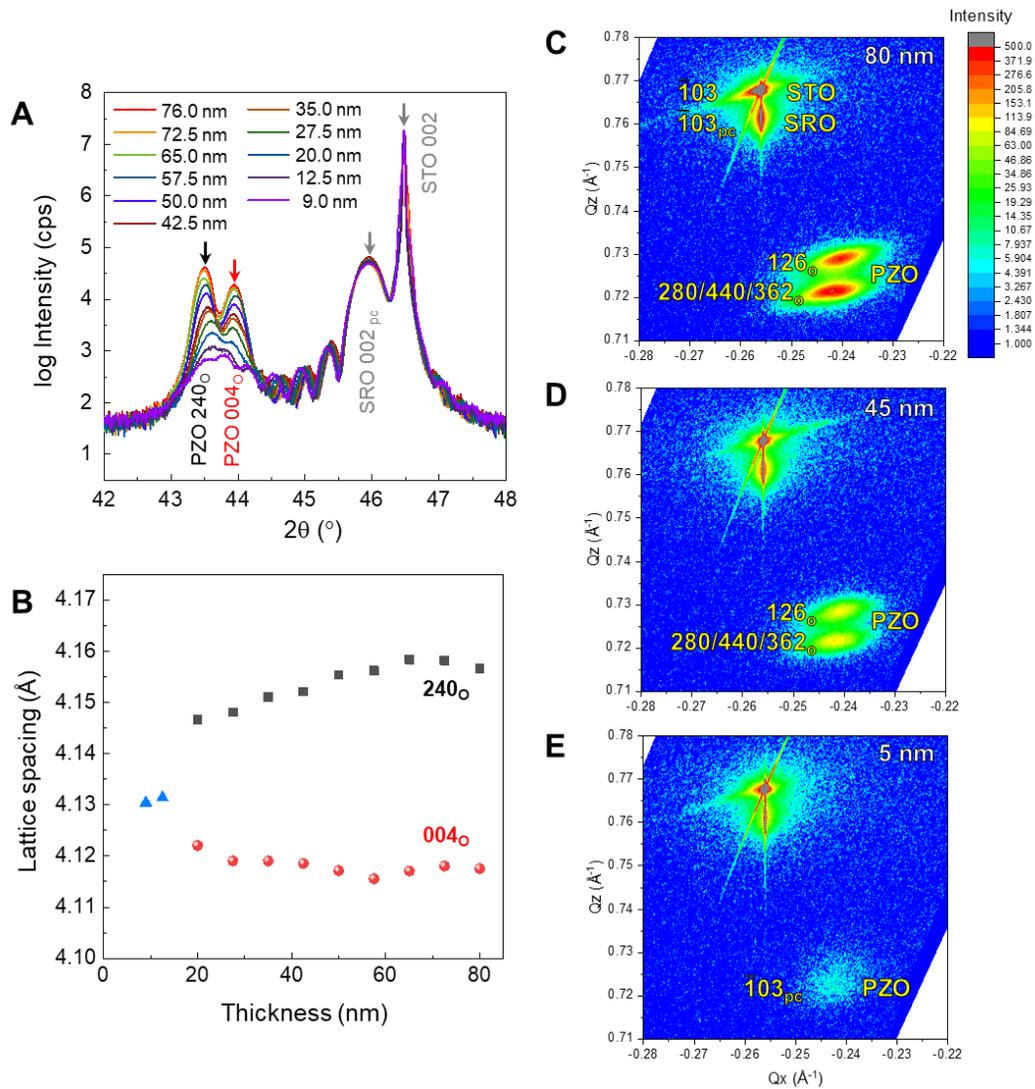

Figure 1. (A) XRD diffraction pattern of a thickness-gradient sample (PZO/SRO/STO) ranging from 5 nm to 80 nm thick with a slope of ~7.5 nm/mm. The sample is 10 mm in length. The pattern exhibits a gradual shift of the PZO peaks, from the two peaks of the orthorhombic AFE phase dominant at higher thickness towards convolution with an emerging third peak with an intermediate lattice parameter, corresponding to the FE phase that becomes dominant in the thin region. (B) The lattice parameter of the PZO shifts with its thickness. (C) − (E) Reciprocal Space Map measurements of a gradient sample (PZO/SRO/STO) with a slope for the PZO of ~7.5 nm/mm, 80 nm, 45 nm and 9 nm thick. The top signal close to STO corresponds to the SRO $(\bar{1}03)_{pc}$ (subscript pc: pseudo-cubic) peak and the low right to the $c$-perpendicular $((362)_O, (280)_O$ and $(440)_O)$ and $c$-parallel $((126)_O)$ domains of PZO.

To determine whether the film exhibits ferroelectric or antiferroelectric characteristics, PFM measurements were carried out. Figures 2(A) and (B) present the PFM phase and amplitude curves for the 80 nm PZO region. The phase curve shows a 180° difference between the positive and negative voltages, attributable to the field-induced ferroelectric states. The amplitude curve exhibits four peaks (indicated by red arrows), and at voltages below the critical voltage for the antiferroelectric to ferroelectric transition, the amplitude signal is very small., suggesting antiferroelectric characteristics.[29,30] The piezoresponse curve was calculated using Amplitude × cos (Phase), and is shown in Figure S1. The double hysteresis loop further suggests the antiferroelectric characteristics. Two additional amplitude curves are displayed in Figure S2, where small amplitudes around zero voltage and the characteristic four amplitude peaks are visible. 2(C) and (D) display PFM phase and amplitude response curves obtained from a region estimated to be 12.5 nm. The 180° phase difference of the hysteresis loop and the butterfly-like amplitude behaviours, indicate the ferroelectric characteristics of the 12.5 nm PZO. The piezoresponse curve of the 12.5 nm PZO is shown in Figure S3. The single hysteresis loop further indicates the ferroelectric characteristics. Phase and amplitude curves of 5 nm PZO are displayed in Figure S4, Though it is noisy, it still shows the ferroelectric characteristics.

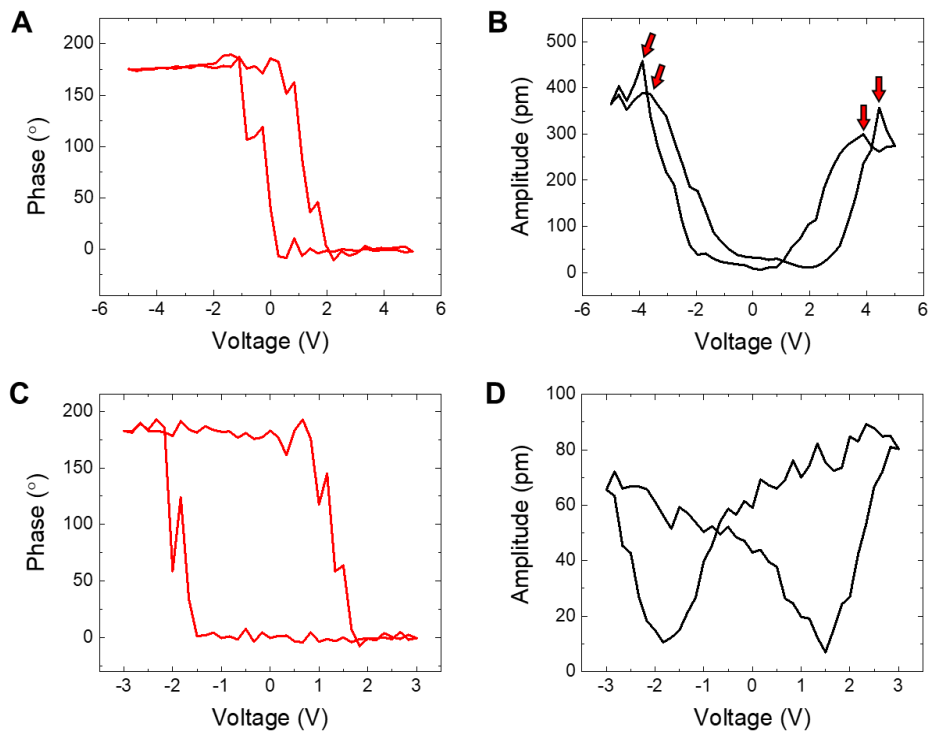

Figure 2. PFM phase (A) and amplitude (B) curves from 80 nm PZO region (voltage on-mode). PFM phase (C) and amplitude (D) curves (voltage off-mode) acquired from the PZO thin film of about 12.5 nm thick.

To investigate the microstructure of PZO at the AFE-FE transition critical thickness, we have prepared a PZO thin film with a constant thickness of 12 nm, grown under the same conditions as the gradient-thickness sample. Figure 3(A) presents the XRD $\theta - 2\theta$ profile of this sample. Peaks corresponding to the STO substrate, SRO buffer layer and PZO are marked with black, blue and purple arrows, respectively. The broad peak for the PZO (002)$_{pc}$ (pc: pseudo-cubic) indicates a variation in the out-of-plane lattice constants. The peak position corresponds to 2θ of 43.6°, and gives rise to an out-of-plane (001)$_{pc}$ lattice spacing of 4.148 Å, consistent with the pseudo-cubic lattice constant ($\sqrt[3]{V}$; $V$, volume) of 4.147Å.[31] The surface morphology and roughness were analyzed using AFM, with results displayed in Figure 3(B), revealing a textured, grainy surface. Figure 3(C) shows a TEM bright field image of the PZO film's cross-section, with clear demarcations of the STO, SRO, and PZO layers. The surface's unevenness is apparent and consistent with the AFM topographical image of the surface shown in Figure 3(B).

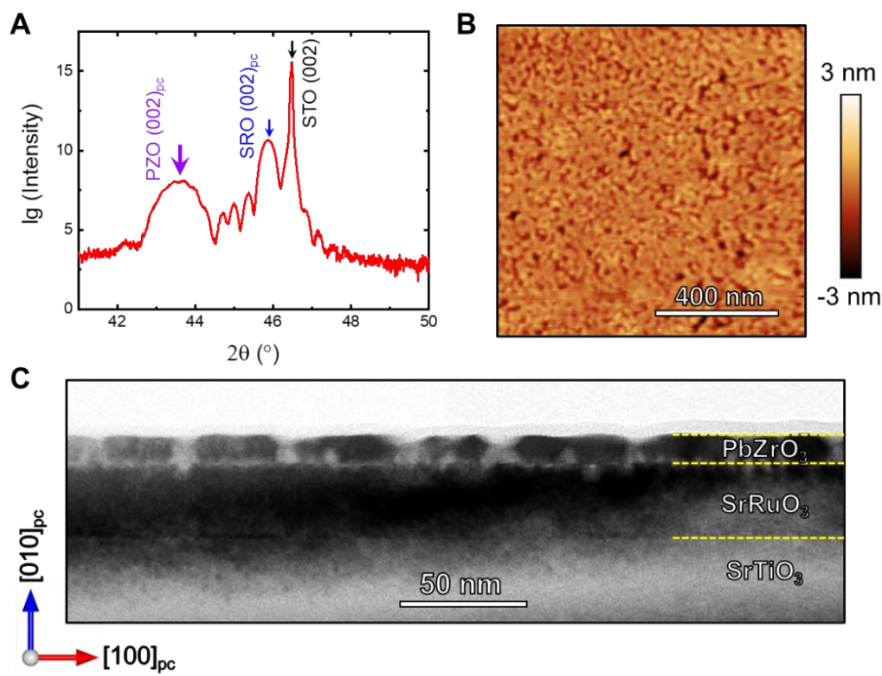

Figure 3. (A) XRD $\theta - 2\theta$ profiles around STO (002) diffraction obtained from a 12 nm PZO thin film that was grown on SRO (about 28 nm) buffered STO (001) substrate. (B) An AFM topographic image showing the surface roughness of the PZO thin film. (C) A TEM bright field image showing the cross-section of the PZO thin film.

Then we further examined the microstructure of the 12 nm PZO thin film using aberration-corrected STEM. A high-resolution STEM-HAADF image showing the PZO (top) and SRO (bottom) layers is shown in Figure 4(A). At the PZO/SRO interface, dislocations with Burgers vector of $a$[100] were revealed most frequently, as outlined using Burgers circuit (white lines) and marked by white arrows, while occasionally ½$a$[101] type dislocations (not shown here) were also observed. This is not quite the same as the primarily ½$a$[101] dislocations reported for an 8 nm thick PZO thin film.[13] The δ$_{Pb}$ map (yellow arrows) was superimposed on the HAADF image, where parallel instead of antiparallel δ$_{Pb}$ is evident, with Pb displacement pointing toward $[10\bar{1}]_{pc}$ projection direction. It suggests the PZO show ferroelectric characteristics, consistent with the PFM measurements in Figures 2 (C) and (D).

The GPA in-plane lattice strain map shown in Figure 4(B) suggests a uniform in-plane strain of about 6% and in-plane lattice constant of 0.414 nm in PZO, in contrast with the antiferroelectric or ferrielectric periodic modulations.[14,32] A Fast Fourier transition (FFT) of the PZO layer is shown in Figure 4(C), from where super-lattice diffraction spots at ½(110) positions were evident as indicated by a yellow arrow, suggesting the formation of superlattice concerning cubic perovskite structure. Such a super-lattice diffraction was considered to be a ferroelectric rhombohedral phase.[13]

To delve into the structure of the ferroelectric phase, iDPC experiments (Figure 4(D)) were carried out, which have been demonstrated to effectively visualize oxygen in perovskite oxides.[33,34] In Figure 4(D), Pb (grey circles), Zr (green circles) and O (red circles) positions are superimposed on the high-resolution iDPC image. The distortion of oxygen octahedra can be determined, as outlined by white lines. A clear difference in the distortion patterns of the neighbouring oxygen octahedra can be identified. This variation is also noticeable in the undulating patterns of oxygen atom chains within the $ZrO_2$ layer – with chains alternating between higher and lower positions horizontally, and left and right positions vertically. These patterns are highlighted by yellow polylines connecting the red circles (O) in Figure 4(D). The oxygen atom columns appear round, indicating that the adjacent oxygen octahedra along the observation direction rotate in the same direction, or in other words, exhibit in-phase rotation (glazer notation of $c^+$). This is different from the antiphase rotation ($c^-$) of the rhombohedral ferroelectric phase with R3c space group.[35] It is also inconsistent with the ferroelectric rhombohedral phase with an R3m space group, which does not show oxygen octahedral distortion.[36] Based on the oxygen position characteristics, a projected 2D unit cell can be determined to be twice that of pseudo-cubic perovskite (as outlined by blue square), and half of the *ab* plane of the orthorhombic antiferroelectric PZO. The determined unit cell is consistent with the superlattice diffraction spots observed in the FFT image in Figure 4(C). If we assume that the differences between the ferroelectric and antiferroelectric phases are solely due to Pb displacement and oxygen octahedral rotation, it would imply that the ferroelectric phase possesses orthorhombic symmetry, akin to antiferroelectric PZO, but with a unit cell size reduced by half. To the best of our knowledge, this phase has not been previously reported in PZO.

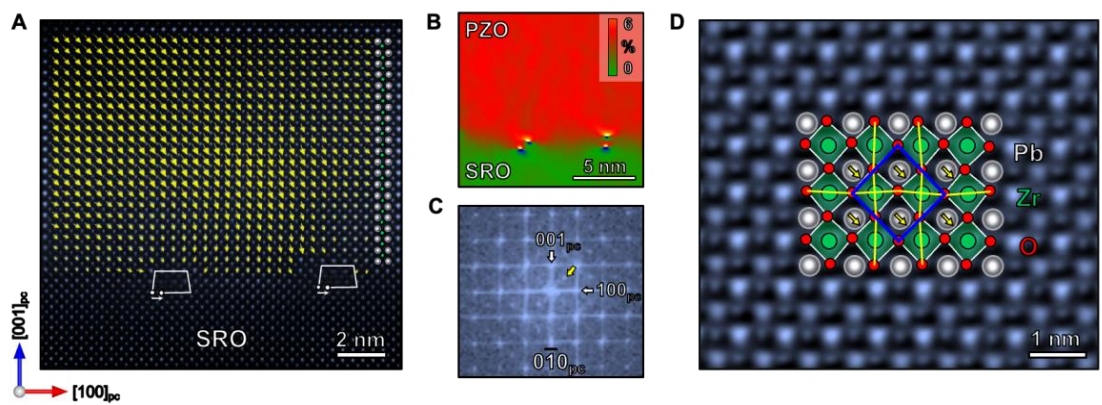

Figure 4. (A) A STEM-HAADF image superimposed by $\delta_{Pb}$ map (yellow arrows) showing the ferroelectric phase observed in PZO ultrathin film (12 nm). (B) GPA of in-plane strain ($\varepsilon_{xx}$) map of HAADF image in (A). (C) FFT of the PZO layer in (A). (D) An iDPC image showing the oxygen positions in the ferroelectric phase observed in PZO ultrathin films. A ferroelectric unit cell is outlined in panel (D).

In addition to the ferroelectric phase, a ferrielectric phase with the dipole of ↑↑·· (with half of the $\delta_{Pb}$ negligible and marked as "·") was also unravelled in the 12 nm thick PZO thin film, as displayed in Figure 5(A) and marked by Region I. The in-plane lattice strain map, generated using GPA and presented in Figure 5(B), reveals a pattern that resembles that of antiferroelectric PZO. This pattern is disrupted in the horizontal direction, as denoted by a green arrow in panel (B). The disruption corresponds to a line where a sudden shift in $\delta_{Pb}$ occurs, also marked by a green arrow in panel (A). The FFT shown in Figure 5(C) exhibits superlattice diffraction spots that correspond to ¼(101)$_{pc}$, indicating that the lattice modulation period is consistent with that of antiferroelectric PZO. Such a ferrielectric state was predicted by theory[37] and observed in doped PZO.[17] In the vicinity of the ferrielectric domain, the direction of lead (Pb) displacement in the ferroelectric regions varies; for instance, on the upper side of the ferrielectric region, the Pb displacement is oriented downwards, avoiding the formation of a sharp head-to-head domain wall.

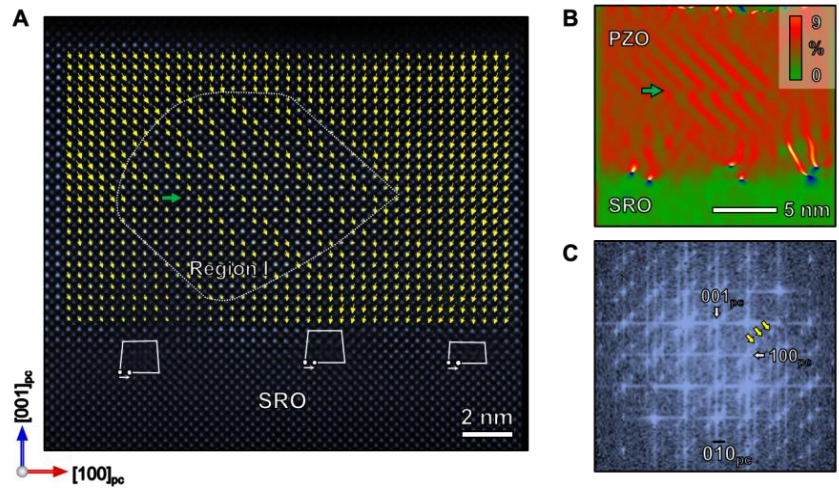

Figure 5. (A) A STEM-HAADF image superimposed by $\delta_{Pb}$ map (yellow arrows) showing the co-existence of ferroelectric and ferrielectric phases in PZO ultrathin films. (B) The GPA in-plane lattice map of (A). (C) Fast Fourier transition of the PZO layer containing region I in (A).

Then we tried to figure out the reason for the formation of such ferroelectric and ferrielectric phases. For PZO, it is well known that the electric field can force the antiferroelectric to ferroelectric transition. Previous first-principles simulation results suggest strain can also cause the AFE to FE phase transition,[38] while previous experimental results on the PZO thin films suggested the strain in the PZO layer has stabilized the ferroelectric rhombohedral phase in the PZO of 8 nm thick.[13] In this work, we have analysed the strain state in the thin film precisely but did not find an obvious difference in the ferroelectric region.

A few factors may contribute to the ferroelectric and ferrielectric phases. The first one is the built-in field caused by the energy-band-mismatch at the PZO/SRO interface. Our PFM results indicate the existence of a built-in bias, as shown in the Amplitude curve of Figure 2B. The bias voltage is of the order of 0.2–0.3V which, divided by the thickness of a 10 nm film translates into a field of 200–300 kV/cm. It is in principle favouring, a ferroelectric transition.

The second one is the interfacial misfit dislocations caused by flexoelectric field. Edge dislocations have been proved to cause a strain gradient along the direction perpendicular to the Burges vector.[33] The flexocoupling coefficient ($f$) of PZO was measured to be around 3 V at room temperature,[39] while the maximum strain gradient ($\nabla S$) is measured to be 0.007733 Å$^{-1}$ for the first unit cell above the dislocation core, it diminishes quickly from the second unit cell. The maximum flexoelectric field $E_f = f * (\nabla S) = 2.3$ MV/cm. This is bigger than the electric field required for antiferroelectric to ferroelectric transition, which is around 400 kV/cm. The flexoelectric field right up the dislocation core points toward the SRO layer. Flexoelectric fields around dislocations therefore seem like an important contribution to the polar tendency of the PZO film.

Surface effects might also play a role in the phase transition. The observation of ferroelectricity in ultrathin antiferroelectric NaNbO$_3$ membranes suggests that distortions / chemical environments beginning at the surface are responsible for driving the ferroelectric phase.[40] Here we have conducted TEM experiments on the surface of PZO thin films. Our TEM results for antiferroelectric thin film (Figure S5) suggest that the top five unit cells (2 nm) are polar, suggesting the surfaces may inherently help stabilize ferroelectricity in very thin films, as predicted by Mani et al.[41] Yet the surface polar layer is only a few unit cells, indicating the surface effect alone may not be sufficient to induce the ferroelectric phase in films that are more than 10 nm thick.

Furthermore, the rotation of oxygen octahedra from the substrate can spread into the thin film, prompting a phase transition. For instance, in thin films of the multiferroic material BiFeO$_3$, oxygen octahedral rotations from the substrate have been shown to transfer to the BiFeO$_3$ layer.[41] In the ferroelectric phase observed in this study, when viewed along the c-axis (as shown in Figure 4), the oxygen octahedra exhibiting in-phase rotation corresponds to the Glazer notation of $c^+$, which is the same as for the SRO electrode. The pattern of oxygen octahedral rotation in SRO is therefore replicated by the PZO, suggesting that octahedral coupling may facilitate the formation of the ferroelectric phase. However, previous studies on BiFeO$_3$ thin films grown on NdGaO$_3$ substrates indicate that only the first three unit cells are influenced by the substrate's oxygen octahedral coupling.[42] Therefore, we believe that coupling oxygen octahedral rotation may help but is unlikely to be the sole factor contributing to the observed ferroelectric phase.

Based on the analysis presented, it is not possible to single out a unique factor – be it the built-in field, the flexoelectric field induced by dislocations, surface effects, or oxygen octahedral rotations – for inducing the ferroelectric phase in the 10 nm films. Rather, it seems likely to be a combination of these influences.

## CONCLUSIONS

PZO thin films with a thickness ranging from 5 nm to 80 nm were fabricated and a transition from AFE to FE phase was confirmed through PFM phase and amplitude curves, suggesting the potential application of ultrathin PZO in memory devices for data storage. Using aberration-corrected STEM-HAADF and iDPC imaging techniques, atomic resolution images were captured from 12nm PZO thin film. It revealed not only Pb and Zr but also O positions, thereby confirming a ferroelectric phase with polarisation along [101]$_{pc}$ projection direction. This is an important result because it indicates that the ferroelectric phase of ultra-thin PbZrO$_3$ is NOT the same as the field-induced ferroelectric rhombohedral phase of bulk PbZrO$_3$. In fact, it is closer to the "ideal" of an antiferroelectric-to-ferroelectric phase transition where the sub-lattice polarisations become parallel to each other by changing the orientation of just one of them, without changing the crystal class.[43] Additionally, a ferrielectric phase characterized by a dipole pattern of ↑↑·· was identified, providing a natural bridging step between the perfectly antipolar and perfectly polar states.

# DECLARATIONS

## Authors' contributions

Experimental design: G.C., Y.L.
Thin film fabrication: Y.L., R.U., J.M.C.R.
XRD measurements and data analysis: J.S., Y.L., D.P., R.U., G.C.
TEM experiment and data analysis: Y.L., R.N., J.C., X.L., G.C.
PFM measurements and data analysis: Y.L., G.C.
Manuscript writing and revision: Y.L. drafted the manuscript; all authors revised the manuscript.
Supervision: G.C., J.A., X.L., J.C.

## Financial support and sponsorship

This project has received funding from the European Union's Horizon 2020 research and innovation program under Grant Agreement No. 766726 (TSAR). Y. L. acknowledges the BIST Postdoctoral Fellowship Programme (PROBIST) funded by the European Union's Horizon 2020 research and innovation programme under the Marie Sklodowska-Curie Grant Agreement No. 754510. D.P. acknowledges funding from 'la Caixa' Foundation fellowship (ID 100010434). X. Z. L. is supported by the Australian Research Council Discovery Project DP190101155. ICN2 acknowledges funding from Generalitat de Catalunya 2021SGR00457. ICN2 is supported by the Severo Ochoa program from Spanish MCIN / AEI (Grant No.: CEX2021-001214-S) and is funded by the CERCA Programme / Generalitat de Catalunya. The authors are grateful for the scientific and technical support from the Australian Centre for Microscopy and Microanalysis (ACMM) as well as the Microscopy Australia node at the University of Sydney.

# Coexistence of ferroelectric and ferrielectric phases in ultrathin antiferroelectric PbZrO$_3$ thin films

Ying Liu[1,2,*], Ranming Niu[2], Roger Uriach[1], David Pesquera[1], José Manuel Caicedo Roque[1], José Santiso[1], Julie M Cairney[2], Xiaozhou Liao[2], Jordi Arbiol[1], Gustau Catalan[1,3,*]

[1]Catalan Institute of Nanoscience and Nanotechnology (ICN2), Campus Universitat Autonoma de Barcelona, Bellaterra 08193, Catalonia, Spain

[2]School of Aerospace, Mechanical and Mechatronic Engineering, The University of Sydney, Sydney, NSW 2008, Australia.

[3]Institut Català de Recerca i Estudis Avançats (ICREA), Pg. Lluís Companys 23, 08010 Barcelona, Catalunya.

**This PDF includes:**

Figure S1 – Figure S5

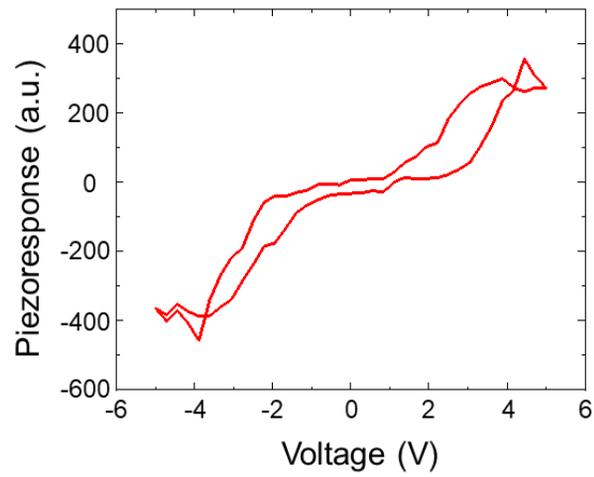

Figure S1. Piezoresponse signal (amplitude multiplied by phase) as a function of tip voltage for PZO thin films of 80 nm. The values for Amplitude and Phase were sourced from Figure 2 in the main text.

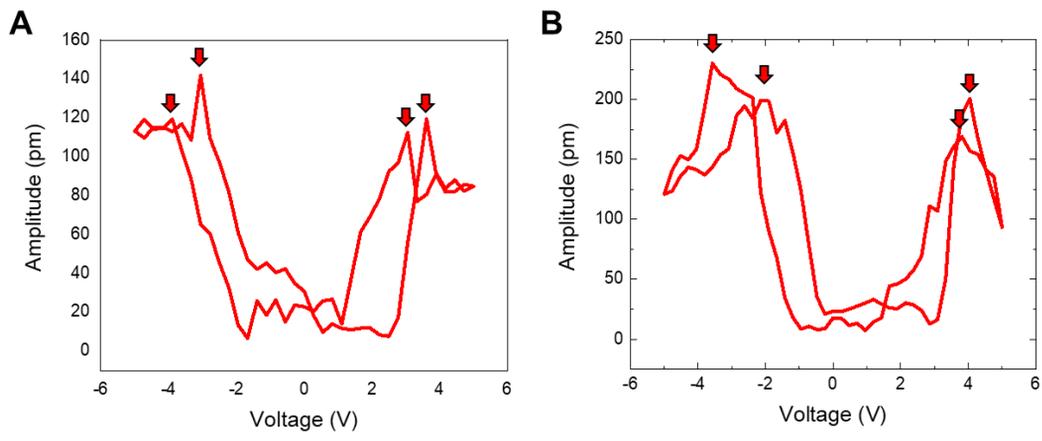

Figure S2. PFM amplitude curves acquired from the 80 nm PZO region showing the antiferroelectric characteristics.

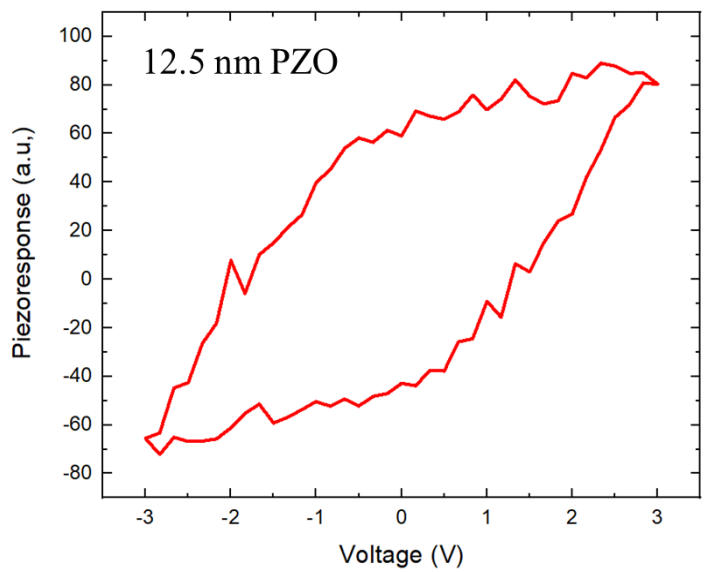

Figure S3. Piezoresponse signal versus voltage for PZO thin films of 12.5 nm thickness, computed as Amplitude × cos (Phase). The values for Amplitude and Phase were sourced from Figure 2 in the main text.

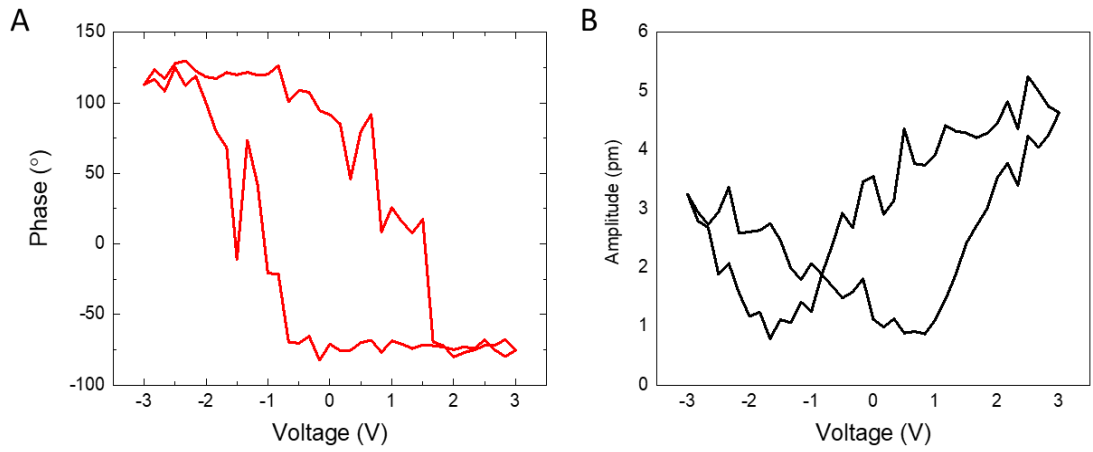

Figure S4. PFM phase (A) and amplitude (B) curves from the 5 nm PZO region showing ferroelectric characteristics.

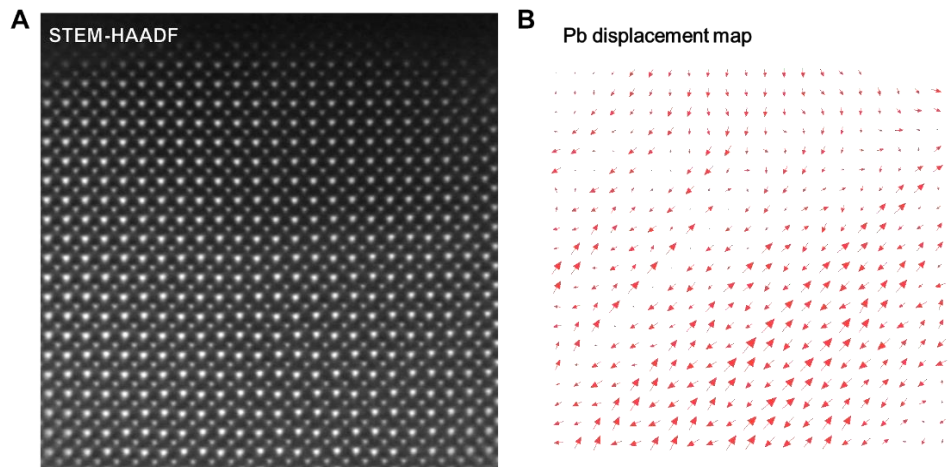

Figure S5. (A) A STEM-HAADF image showing the atomic structure of the PZO single crystal surface. (B) Pb displacement map extracted from (A).